\documentclass[square]{ws-procs9x6}  

\usepackage{amssymb,amsmath,float,nicefrac,euscript}

\newcommand{\remove}[1]{}

\def\wt{\qopname\relax{no}{w}}
\def\proj{\qopname\relax{no}{proj}}

\newcommand\nc\newcommand
\nc\bfa{{\boldsymbol a}}\nc\bfA{{\mathbf A}}\nc\cA{{\mathcal A}} 
\nc\bfb{{\boldsymbol b}}\nc\bfB{{\mathbf B}}\nc\cB{{\mathcal B}} 
\nc\bfc{{\boldsymbol
c}}\nc\bfC{{\mathbf C}}\nc\cC{{\mathcal C}} \nc\bfd{{\boldsymbol
d}}\nc\bfD{{\mathbf D}}\nc\cD{{\mathcal D}} \nc\bfe{{\boldsymbol
e}}\nc\bfE{{\mathbf E}}\nc\cE{{\mathcal E}} \nc\bff{{\boldsymbol
f}}\nc\bfF{{\mathbf F}}\nc\cF{{\mathcal F}} \nc\bfg{{\boldsymbol
g}}\nc\bfG{{\mathbf G}}\nc\cG{{\mathcal G}} \nc\bfh{{\boldsymbol
h}}\nc\bfH{{\mathbf H}}\nc\cH{{\mathcal H}} \nc\bfi{{\boldsymbol
i}}\nc\bfI{{\mathbf I}}\nc\cI{{\mathcal I}} \nc\bfj{{\boldsymbol
j}}\nc\bfJ{{\mathbf J}}\nc\cJ{{\mathcal J}} \nc\bfk{{\boldsymbol
k}}\nc\bfK{{\mathbf K}}\nc\cK{{\mathcal K}} \nc\bfl{{\boldsymbol
l}}\nc\bfL{{\mathbf L}}\nc\cL{{\mathcal L}} \nc\bfm{{\boldsymbol
m}}\nc\bfM{{\mathbf M}}\nc\cM{{\mathcal M}} \nc\bfn{{\boldsymbol
n}}\nc\bfN{{\mathbf N}}\nc\cN{{\mathcal N}} \nc\bfo{{\boldsymbol
o}}\nc\bfO{{\mathbf O}}\nc\cO{{\mathcal O}} \nc\bfp{{\boldsymbol
p}}\nc\bfP{{\mathbf P}}\nc\cP{{\mathcal P}} \nc\bfq{{\boldsymbol
q}}\nc\bfQ{{\mathbf Q}}\nc\cQ{{\mathcal Q}} \nc\bfr{{\boldsymbol
r}}\nc\bfR{{\mathbf R}}\nc\cR{{\mathcal R}} \nc\bfs{{\boldsymbol
s}}\nc\bfS{{\mathbf S}}\nc\cS{{\mathcal S}} \nc\bft{{\boldsymbol
t}}\nc\bfT{{\mathbf T}}\nc\cT{{\mathcal T}} \nc\bfu{{\boldsymbol
u}}\nc\bfU{{\mathbf U}}\nc\cU{{\mathcal U}} \nc\bfv{{\boldsymbol
v}}\nc\bfV{{\mathbf V}}\nc\cV{{\mathcal V}} \nc\bfw{{\boldsymbol
w}}\nc\bfW{{\mathbf W}}\nc\cW{{\mathcal W}} \nc\bfx{{\boldsymbol
x}}\nc\bfX{{\mathbf X}}\nc\cX{{\mathcal X}} \nc\bfy{{\boldsymbol
y}}\nc\bfY{{\mathbf Y}}\nc\cY{{\mathcal Y}} \nc\bfz{{\boldsymbol
z}}\nc\bfZ{{\mathbf Z}}\nc\cZ{{\mathcal Z}} \nc\od{{\bar d}}\nc\ow{{\bar
w}}\nc\odelta{{\bar\delta}} \nc\ox{{\bar x}}\nc\oy{{\bar y}}\nc\ou{{\bar u}}
\nc\oh{{\bar h}}

\newcommand\reals{{\mathbb R}}\nc{\nd}{\noindent}

\newcommand\bP{{\mathbb P}}
\newcommand\integers{{\mathbb Z}}
\nc\dgv{\delta_{\text{\rm GV}}} \nc\rcrit{R_{\text{\rm crit}}}
\nc\Esp{E_{\text{\rm sp}}}
\renewcommand\epsilon{\varepsilon}

\newcommand{\half}{\nicefrac12}
\newcommand{\gX}{\mathfrak X}
\newcommand{\beeq}{\begin{eqnarray*}}
\newcommand{\eneq}{\end{eqnarray*}}

\begin{document}

\title{A functional view of upper bounds on codes}

\author{Alexander Barg$^{1,2}$ and Dmitry Nogin$^2$}

\address{$^1$ Department of ECE/ISR\\
University of Maryland\\College Park MD 20742, USA\\[3mm]
$^2$Dobrushin Math. Laboratory\\
Institute for Information Transmission Problems\\
Bol'shoj Karetnyj 19, Moscow 101447, Russia\\[2mm]
E-mails: \{abarg@umd.edu, nogin@iitp.ru\}}

\begin{abstract}
Functional and linear-algebraic approaches to the Delsarte problem
of upper bounds on codes are discussed.
We show that Christoffel-Darboux kernels and Levenshtein polynomials
related to them arise as stationary points of the moment functionals of some
distributions.
We also show that they can be derived as eigenfunctions of the Jacobi
operator.
\end{abstract}


\bodymatter

\section{Introduction}\label{aba:sec1}
In the problem of bounding the size of codes in compact homogeneous
spaces, Delsarte's polynomial method gives rise to the most powerful
universal bounds on codes. Many overviews of the method exist in 
the literature; see for instance Levenshtein \cite{lev98}. 
In this note, which extends our previous work \cite{bar06a} 
we develop a functional perspective of this method and give some
examples. We also discuss another version of the functional approach,
a linear algebraic method for the construction of polynomials for Delsarte's
problem. Our main results are new constructions of Levenshtein's polynomials.

Let $\gX$ be a compact metric space with distance function $\tau$ whose 
isometry group $G$ acts
transitively on it. The zonal polynomials associated with this action
give rise to a family of orthogonal polynomials $\cP(\gX)=\{P_\kappa\}$ where
$\kappa=0,1,\dots$ is the total degree.
These polynomials are univariate if $G$ acts on $\gX$ doubly transitively 
(the well-known examples include the Hamming and Johnson graphs, their 
$q$-analogs and other $Q$-polynomial distance-regular graphs; the sphere 
$S^{d-1} \in\reals^d$) and are multivariate otherwise. 

First consider the univariate case. Then for any given value of the
degree $\kappa=i$ the family $\cP(\gX)$ contains only one polynomial of
degree $i$, denoted below by $P_i.$
Suppose that the distance on $\gX$ is measured in such a way that 
$\tau(x,x)=1$ and the diameter of $\gX$ equals $-1$ (to accomplish this, a 
change of variable is made in the natural distance function on $\gX$).
We refer to the model case of $\gX=S^{d-1}$ although the arguments below
apply to all spaces $\gX$ with the above properties.
Let $\langle f,g\rangle=\int_{-1}^1 fgd\mu$ be the inner product in 
$L_2([-1,1],d\mu)$ where $d\mu(x)$ is a distribution on $[-1,1]$ induced by 
a $G$-invariant measure on $\gX.$ Let $\cF(\cdot)\triangleq\langle \cdot,1\rangle$
be the moment functional with respect to $d\mu.$
We assume that this distribution is normalized, i.e., that 
$\cF(1)=1.$

Let $C$ be a code, i.e., a finite collection of points in $\gX$.
By Delsarte's theorem, the size of the code $C$
whose distances take values in $[-1,s]$ is bounded above by 
  \begin{equation}\label{eq:cond1}
       |C|\le \inf_{f\in \Phi} f(1)/\hat f_0,
  \end{equation}
where 
  \begin{equation}\label{eq:conditions}
     \Phi=\Phi(s)\triangleq\{f: f(x)\le 0, x\in [-1,s];\quad
   \hat f_0>0, \,\;\hat f_i\ge 0, i=1,2,\dots\}
  \end{equation}
is the cone of positive semidefinite functions that are nonpositive on 
$[-1,s]$ (here $\hat f_i=\langle f,P_i \rangle/\langle P_i,P_i\rangle$ 
are the Fourier coefficients of $f$).

\section{Functional approach}
The choice of polynomials for problem (\ref{eq:cond1})-(\ref{eq:conditions})
was studied extensively in the works of Levenshtein \cite{lev78,lev83a,lev92}. 
In this section we give a new construction of his polynomials and their
simplified versions.

\subsection{Notation.} 
Let $V$ be the space of real square-integrable functions on $[-1,1]$ and
let $V_k$ be the space of polynomials of degree $k$ or less.
Let $p_i=P_i/\langle P_i,P_i\rangle, i=0,1,\dots$ be the normalized 
polynomials. The polynomials $\{p_i\}$ satisfy a three-term recurrence
of the form
    \begin{align}\label{eq:3term}
      xp_i=a_ip_{i+1}&+b_ip_i+a_{i-1}p_{i-1},\\ & i=1,2,\dots;p_{-1}=0,
p_0=1;\; a_{-1}=0.\nonumber
    \end{align}
In other words, the matrix of the operator $x:V\to V$ (multiplication
by the argument) in the orthonormal basis is a semi-infinite symmetric 
tridiagonal matrix, called the Jacobi matrix. 
Let $X_k=E_k\circ x$ where $E_k=\proj_{V\to V_k}$, and let $J_k$ be 
the $(k+1)\times(k+1)$ submatrix of $J$,
  $$J_k=\begin{bmatrix}b_0 &a_0 & 0& 0&\dots&0\\
       a_0 &b_1 &a_1& 0&\dots&0\\
        0&a_1&b_2&a_2&\dots&0\\
       \dots &\dots &\dots&\dots &\dots&a_{k-1}\\
       0&0 &\dots&\dots &a_{k-1} &b_k
     \end{bmatrix}.
  $$
\remove{  $$
    J_k=\begin{pmatrix}\ast &a_0& a_1&\dots&a_{k-2}&a_{k-1}\\
         b_0&b_1&b_2&\dots&b_{k-1}&b_k\\
        a_0&a_1&a_2&\dots&a_{k-1}&\ast
     \end{pmatrix}
  $$
(we show only the diagonals, the $\ast$'s compensate for their unequal length).}
\begin{example}\label{example} $(a)$ For instance, let $\gX$ be the binary $n$-dimensional Hamming
space. Then
   $p_i(x)=\tilde k_i(\nicefrac n2(1-x)),$
where $\tilde k_i(z)$ is the normalized Krawtchouk polynomial. The 
polynomials $p_i(x)$ are orthogonal on the finite set of points
  $
    \{x_j=1-(2j/n),j=0,1,\dots,n\}$
with weight $w(x_j)=\binom n{j}2^{-n}$ and have unit norm.
In this case, 
   \begin{equation}\label{eq:ai}
    a_i=(1/n)\sqrt{(n-i)(i+1)}, \;b_i=0, \quad 0\le i\le n.
   \end{equation}
$(b)$ Let $\gX$ be the unit sphere in $d$ dimensions. Then $p_i(x)$ are the
normalized Gegenbauer polynomials; in this case
  $$
   a_i=\sqrt{\frac{(n-i+2)(i+1)}{(n+2i)(n+2i-2)}}, \;b_i=0,\quad i=0,1,\dots.
  $$
\end{example}

\medskip
It is well known \cite[p.243]{and99} that for $k\ge 1$ the spectrum of 
$X_k$ coincides with the set 
${\EuScript X}_{k+1}=\{x_{k+1,1},\dots,x_{k+1,k+1}\}$ of 
zeros of $p_{k+1}.$ Below we denote the largest of these zeros by $x_{k+1}.$ 
Let
   \begin{equation}\label{eq:Kk}
     K_k(x,s)\triangleq \sum_{i=0}^k p_i(s)p_i(x)
   \end{equation}
be the $k$-th reproducing kernel.
By the Christoffel-Darboux formula, 
    \begin{equation}\label{eq:cd}
      (x-s)K_k(x,s)=a_k(p_{k+1}(x)p_k(s)-p_{k+1}(s)p_k(x)).
    \end{equation}
In particular, if $s \in {\EuScript X}_{k+1}$ then $X_k K_k(x,s)=s K_k(x,s).$
Note that $K_k(x,y)$ acts on $V_k$ as a delta-function at $y$:
    \begin{equation}\label{eq:delta}
      \langle K_k(\cdot,y),f(\cdot)\rangle=f(y).
    \end{equation}

\subsection{Construction of polynomials.}
Without loss of generality let us assume that $f(1)=1.$ Then (\ref{eq:cond1})
is equivalent to the problem
    $$
     \sup\{\cF(f), f\in \Phi\}.
    $$
Let us restrict the class of functions to $V_n.$ 
By the Markov-Lucacs theorem \cite[Thm.~6.4]{kre77}, a polynomial $f(x)$
that is nonpositive on $[-1,s]$ can be written in the form
   $$
     f_n(x)=(x-s)g^2-(x+1)\phi_1^2 \quad\text{or}\quad f_n(x)=(x+1)(x-s)g^2-
     \phi_2^2
   $$
according as its degree $n=2k+1$ or $2k+2$ is odd or even. 
Here $g,\phi_1\in V_k, 
\phi_2\in V_{k+1}$ are some polynomials.
Below the negative terms will
be discarded. We use a generic notation $c$ for multiplicative constants
chosen to fulfill the condition $f(1)=1.$

\subsubsection{The MRRW polynomial.}
Restricting our attention to odd degrees $n=2k+1$, let us seek
$f(x)$ in the form $(x-s)g^2$.  Let us write the Taylor expansion
of $\cF$ in the ``neighborhood'' of $g$.
Let $h\in V_k$ be a function that satisfies $\|h\|\le \epsilon$ for a small
positive $\epsilon$ and the condition $h(1)=0.$  We obtain
  \begin{equation*}\label{eq:Taylor}
   \cF((x-s)(g+h)^2)=\cF((x-s)g^2)+
  \langle(x-s)(g+h),g+h\rangle-\langle(x-s)g,g\rangle\end{equation*}
  $$   =\cF(f)+ \cF'(h)+\half\langle\cF'' h,h\rangle,
  $$
where  $\cF'=2(x-s)g, \cF''=2(x-s)$ are the Fr{\'e}chet derivatives 
of $\cF.$  
This relation shows that for $f$ to be a stationary point 
of $\cF$, the function $g$ should satisfy $d\cF=2\langle g,(x-s)h\rangle=0$ 
for any function $h$ with the above properties. First assume that $s=x_{k+1}.$
Then by (\ref{eq:cd}), a stationary point of $\cF$
is given by $g=K_k(x,s),$ and we obtain $f$ in the form
   $$
     f_n(x)=c(x-s)(K_k(x,s))^2.
   $$
Since $\hat f_0=0,$ conditions (\ref{eq:conditions}) are not satisfied;
however, it is easy to check that they are satisfied if 
$x_{k}<s<x_{k+1}.$ For all such $s,$ the polynomial $f_n$ is a valid choice
for problem (\ref{eq:cond1}), yielding
   \begin{equation}\label{eq:mrrw}
    |C|\le -\frac{1-s}{a_kp_{k+1}(s)p_k(s)}K_k^2(1,s).
   \end{equation}
The polynomial $f_n$ was used by McEliece {\em et al.} \cite{mce77a}
and Kabatiansky and Levenshtein \cite{kab78} to derive their well known 
upper bounds on codes.

\subsubsection{Levenshtein polynomials, $n=2k+1.$}
So far in our optimization we did not use the condition
$h(1)=0.$ To use it, let us write $h=(1-x)h_1, h_1\in V_{k-1}$ and
repeat the above calculation. We find that stationary points of $\cF$
should satisfy
   $$
     d\cF^{(-)}=2\langle(x-s)g,(1-x)h_1\rangle=0,
   $$
where $\cF^{(-)}(\,.\,)=\int .\,(1-x)d\mu$ is the moment functional 
with respect to the
distribution $d\mu^{(-)}(x)=(1-x)d\mu(x).$ 
A stationary point of $\cF^{(-)}$ is given by the reproducing kernel 
$K_k^{-}(x,s)$ {\sl with respect to this distribution}:
   \begin{equation}\label{eq:lv1}
     K_k^{-}(x,s)=\sum_{i=0}^k p_i^{-}(s)p_i^{-}(x),
   \end{equation}
where $\{p_i^-(x),i=0,1,\dots\}$ is the corresponding orthonormal system. 
To find the polynomials $p_i^-(x)$ observe that
   $$
    \cF^{(-)}(p_i^-p_j^-)=\cF(p_i^-(x)p_j^-(x)(1-x))= \delta_{i,j}
   $$
is satisfied for $p_i^{-}(x)=K_i(1,x)/(a_ip_{i+1}(1)p_i(1))^{\half}.$ 
Indeed, if $j<i$ then the function $(1-x)K_i(1,x)$ is in the subspace 
spanned by $p_{i+1},p_i$ and thus is orthogonal to $K_j(1,x).$
To conclude, the function sought can be taken in the form
  $$
   f_n^-(x)=c(x-s)(K_k^-(x,s))^2.
  $$

\vspace*{-7mm}

\subsubsection{Levenshtein polynomials, $n=2k+2.$}
In this case we seek the polynomial in the form $f_n=(x-s)(x+1)g^2.$
The necessary condition for the stationary point takes the form
$\cF^\pm((x-s)gh)\triangleq\langle (x-s)(1-x^2) g,h\rangle=0$. 
From this, $g=K_{k}^\pm(x,s)$
where the kernel $K_{k}^\pm$ is taken with respect to the distribution
$d\mu^{(\pm)}(x)=(1+x)(1-x) d\mu(x).$ The corresponding orthogonal 
polynomials $p^{\pm}_i(x)$ are also easily found: up to normalization
they are equal
      $$
      p^\pm_i(x)=K_i(x,-1)p_{i+1}(1)-K_i(x,1)p_{i+1}(-1).
      $$
Then 
   $$
      f_n^\pm(x)=c(x-s)(x+1)(K_{k}^\pm(x,s))^2.
   $$
Let $x_k^-$ ($x_k^\pm$) be the largest root of 
$p_k^-(x)$ (resp. of $p_k^\pm(x)$). Then $f_{2k+1}^-(x)\in \Phi$ 
if $x_k^{\pm} \le s\le x_{k+1}^-$ and $f_{2k+2}^{\pm}(x)\in \Phi$
if $x_{k+1}^-<s<x_{k+1}^\pm.$

{\em Remarks.}

1. The polynomials $f_n^-,f_n^\pm$ were constructed and applied to
coding theory by Levenshtein \cite{lev78,lev83a,lev92}. 
Polynomials closely related to them were studied in a more general context 
in the works of M.~G.~Krein {\em et al.}; see Krein and Nudelman \cite{kre77}. 
The orthogonal systems $\{p^-_i\},\{p^\pm_i\}$ are sometimes 
called {\sl adjacent polynomials} of the original system $\{p_i\}.$ 

2. The stationary points found above are not true extremums
because the
second differential of the functionals $\cF, \cF^{(-)}, \cF^{(\pm)}$ 
is indefinite: for instance, $d^2\cF(g)=2\langle (x-s)h,h\rangle$.
Nevertheless, the polynomials $f_n^-,f_n^\pm$ have been proved \cite{sid80}
to be optimal
in the following sense: for any $n\ge 1$ and all $f\in \Phi, \deg f\le n$
   $$
    \cF(f_n)\ge \cF(f).
   $$

3. Asymptotic bounds derived from (\ref{eq:cond1}) 
relying upon the polynomials $f_n,f^-_n,f^\pm_n$ coincide. For the finite
values of the parameters, better bounds are obtained from $f^-_n,f^\pm_n.$

\subsection*{3. Spectral method}
This section is devoted to a different way of constructing polynomials
for the Delsarte problem. The ideas discussed below originate in the 
work of C.~Bachoc \cite{bac06}. They were elaborated upon in an earlier
work of the authors \cite{bar06a}.

We develop the remark made after (\ref{eq:cd}), namely that for any $i\ge 1,$
$K_k(x,x_{k+1,i})$ is an eigenfunction of the Jacobi operator $X_k.$ 
Since $K_k(x,s)$ is a good choice for the polynomial in Delsarte's problem,
it is possible to construct polynomials as
eigenvectors of $X_k$ as opposed to the analytic arguments discussed
above. In particular, 
$K_k(x,s)$ arises as an eigenfunction of the operator
$T_k=T_k(s)$ defined by
  \begin{align*}
      T_k: &V_k\to V_k\\
        &\phi\mapsto X_k\phi +\rho_k         \hat\phi_kp_k
  \end{align*}
where $\rho_k=a_k p_{k+1}(s)/p_k(s).$ 
Indeed, using (\ref{eq:Kk}) and (\ref{eq:cd}) we obtain
  $$
   (T_k-s)K_k(x,s)=(X_k-s)K_k(x,s)+a_k p_{k+1}(s)p_k(x)=0.
  $$
On account of earlier arguments 
we should choose the polynomial for problem (\ref{eq:cond1}) in the
form $F(x)=(x-s)f^2(x)$ where $f(x)=f(x,s)$ is an eigenfunction of $T_k.$
 The positive definiteness condition of $f$  
can be proved using the Perron-Frobenius theorem; for this we must take
$f$ to be the eigenfunction that corresponds to the {\em largest} eigenvalue
of $T_k.$ This condition defines the range of code 
distances $s$ in which the method is applicable.

A variant of this calculation was performed in \cite{bar06a} to which
we refer for details. The difference between \cite{bar06a} and
the argument above is that there we took $\rho_k=a_kp_{k+1}(1)/p_k(1).$
This has the advantage of defining $T_k$ independently of $s$ but leads to
a bound of the form
  \begin{equation}\label{eq:bound}
      |C|\le \frac{4a_kp_{k+1}(1)p_k(1)}{1-\lambda_k}
   \end{equation}
which is generally somewhat weaker than (\ref{eq:mrrw}). 
Using the function $F$ defined above we can improve this to recover
the estimate (\ref{eq:mrrw}).

We note that this argument does not depend on the 
choice of the functional space; in particular, the kernels $K_k^-, K_k^\pm$ 
arise if the operator $X_k$ is written with respect to the basis of the corresponding adjacent polynomials 
($\{p^-_i\}$ or $\{p^\pm_i\}$) and their generating distribution. 
To conclude, Levenshtein's polynomials and bounds on codes can be
derived within the framework of the spectral method.

\begin{example} Consider again Example \ref{example}$(a)$. The adjacent
polynomials up to a constant factor that does not depend on $i$
are given by \cite[p.81]{lev83a}
  $$
    p_i^-(x)=\tilde k^{(n-1)}_i (z), \;
     p_i^\pm(x)=\tilde k^{(n-2)}_i (z)  \quad\text{ for }z=\frac n2(1-x)-1,
  $$
where $\tilde k^{(n-1)}_i(z)$ for instance denotes the degree-$i$
normalized 
Krawtchouk polynomial associated with the $(n-1)$-dimensional Hamming space. 
The Jacobi matrix $J_k$ for the basis ${p_i^-}$ can be computed from 
(\ref{eq:ai}) as follows. Since
  $$
   x p_i^-(x)=\Big(1-\frac 2n(z+1)\Big)\tilde k_i^{(n-1)}(z), 
   $$           
we find that the coefficients of three-term recurrence for the family
$\{p_i^-\}$ are 
   $$
   a_i=(1/n)\sqrt{(n-k-1)(k+1)}, \;b_i=-1/n, \;\;i=0,1,\dots.
   $$
Constructing the operator $T_k$ as described above, we obtain $K_k^-(x,s)$
as its eigenfunction. A similar construction can be pursued for the 
function $K_k^\pm.$
\end{example} 

The approach outlined above has two advantages. First, it enables one to 
obtain simple estimates of the largest eigenvalue of $X_k$ which is 
important in verifying the condition $f(x)\le 0, x\in[-1,s].$
The second advantage is a more substantial one: this method can
be extended to the case of {\sl multivariate zonal polynomials}
when the analytic alternative is not readily available.
This case arises when the space $\gX$ is homogeneous but not 2-point
homogeneous. Worked examples include the real Grassmann manifold $G_{k,n}$
(\cite{bac06}; the $P_i$ are given by the generalized $k$-variate Jacobi
polynomials) and the so-called ordered Hamming space \cite{bar07c}. 
We provide a few more details on the latter
case in order to illustrate the general method.

Let $\cQ$ be a finite alphabet of size $q.$ Consider the set $\cQ^{r,n}$ of 
vectors of dimension $rn$ over $\cQ$. A vector $\bfx$ will be written as
a concatenation of $n$ blocks of length $r$ each, 
$\bfx=\{x_{11},\dots,x_{1r};\dots;x_{n1},\dots,x_{nr}\}.$ 
For a given vector $\bfx$ let $e_i, i=1,\dots,r$ be the number of $r$-blocks
of $\bfx$ whose rightmost nonzero entry is in the $i$th position counting
from the beginning of the block. The $r$-vector $e=(e_1,\dots,e_r)$ will
be called the {\em shape} of $\bfx$. 
A shape vector $e=(e_1,\dots,e_r)$ 
defines a partition of a number $N\le n$ into a sum of $r$
parts.  Let $e_0=n-\sum_i e_i.$
Let 
$\Delta_{r,n}=\{e\in (\integers_+\cup\{0\})^r: \sum_ie_i\le n\}$ 
be the set of all such partitions.
The zonal polynomials associated to $\cQ^{r,n}$ are $r$-variate polynomials
$P_f(e), f,e\in \Delta_{r,n}$ of degree $\kappa=\sum_i f_i$. They are
orthogonal on $\Delta_{r,n}$ according to the following inner product
   $$
     \sum_{e\in \Delta_{r,n}} P_f(e)P_g(e)w(e)
            =0   \quad (f\ne g).
  $$
The weight in this relation is given by the multinomial probability
distribution
  $$
    w(e_1,\dots, e_r)=n!\prod_{i=0}^r \frac{p_i^{e_i}}{e_i!} \qquad
(p_i=q^{i-r-1}(q-1), i=1,\dots, r; p_0=q^{-r}),
  $$
so the polynomials $P_f(e)$ form a particular case of 
{\em r-variate Krawtchouk polynomials}. 

Let $\bfx\in \cQ^{r,n}$ be a vector of shape $e$.
Define a norm on $\cQ^{r,n}$ by setting $\wt(\bfx)=\sum_i ie_i$
and let $d_r(\bfx,\bfy)=\wt(\bfx-\bfy)$ be the ordered Hamming metric
(known also as the Niederreiter-Rosenbloom-Tsfasman metric). We note that
in the multivariate case there is no direct link between the
variables and the metric. For instance, for the space $\cQ^{r,n}$
the polynomials (as well as relations in the corresponding association scheme)
are naturally indexed by shape vectors $e$ while the weight is some function 
$e.$ 

The Delsarte theorem in this case takes the following form: {\em  
The size of an $(n,M,d)$ code $C\subset \cQ^{r,n}$ is bounded above by}
  $    M\le \inf_{f\in \Phi}f(0)/f_0,  $ where
\begin{align*}
   \Phi=\{f(x)=f(x_1,\dots,&x_r)=
     f_0+\sum_{e \ne 0} f_e P_e(x):  f_0>0,f_e\ge 0\;(e\ne 0); \\[-3mm]
    &f(e)\le 0 \;\;\forall e \text{ \rm s.t. } \sum_{i=1}^r ie_i\le d\}
\end{align*}
The argument for the univariate case given in this section
can be repeated once we establish a three-term relation for the 
polynomials $P_f(e).$ Let $\bP_\kappa$
be the column vector of the normalized 
polynomials $P_f$ ordered lexicographically
with respect to all $f$ that satisfy $\sum_i f_i=\kappa$ and let $F(e)$
be a suitably chosen linear polynomial. Then
   $$
   F(e)\bP_\kappa(e)=A_\kappa \bP_{\kappa+1}(e)+B_\kappa\bP_{\kappa}(e)
      +A_{\kappa-1}^T \bP_{\kappa-1}(e)
   $$
where  $A_\kappa,B_\kappa$ are matrices of order 
$\binom{\kappa+r-1}{r-1}\times\binom{\kappa+s+r-1} {r-1}$ and $s=1,0,$ 
respectively.
The elements of these matrices can be computed explicitly from combinatorial
considerations. This gives an explicit form of the operator
$S_\kappa=E_{\kappa}\circ F(e)$ in the orthonormal basis. Relying on this,
it is possible to derive a bound on codes in the NRT space of the form
(\ref{eq:bound}) and perform explicit calculations, both in the case of
finite parameters and for asymptotics.
The full details of the calculations are given in \cite{bar07c}.

\nd {\bf Acknowledgments:} 
The research of A. Barg is supported in part
by NSF grants
CCF0515124 and CCF0635271, and by NSA grant H98230-06-1-0044. The research
of D. Nogin is supported in part by Russian Foundation for Basic
Research through grants RFBR 06-01-72550-CNRS and RFBR 06-01-72004-MST.
Parts of this research were presented at the International
Workshop on Coding and Cryptology, The Wuyi Mountain, Fujian, China, June 11-15, 2007, and COE Conference on the 
Development of Dynamic Mathematics with High Functionality
(DMHF2007), Fukuoka, Japan, October 1-4, 2007.


\begin{thebibliography}{12}
\providecommand{\natexlab}[1]{#1}
\providecommand{\url}[1]{\texttt{#1}}
\expandafter\ifx\csname urlstyle\endcsname\relax
  \providecommand{\doi}[1]{doi: #1}\else
  \providecommand{\doi}{doi: \begingroup \urlstyle{rm}\Url}\fi

\bibitem{lev98}
V.~I. Levenshtein.
\newblock Universal bounds for codes and designs.
\newblock In V.~Pless and W.~C. Huffman, Eds., \emph{Handbook of Coding Theory},
  vol.~1, pp. 499--648. Elsevier Science, Amsterdam,  (1998).


\bibitem{bar06a}
A.~Barg and D.~Nogin, Spectral approach to linear programming bounds on codes,
  \emph{Problems of Information Transmission}. {\bf 42}, \penalty0 12--25,
  (2006).

\bibitem{bar07c}
A.~Barg and P.~Purkayastha, Bounds on ordered codes and orthogonal arrays.
\newblock arxiv:CS/0702033.

\bibitem{lev78}
V.~I. Levenshtein.
\newblock On choosing polynomials to obtain bounds in packing problems.
\newblock In \emph{Proc. 7th All-Union Conf. Coding Theory and Information
  Transmission,Part 2}, pp. 103--108, Moscow, Vilnius,  (1978).
\newblock (In Russian).

\bibitem{lev83a}
V.~I. Levenshtein, Bounds for packings of metric spaces and some of their
  applications, \emph{Problemy Kibernet.} {\bf 40}, \penalty0 43--110 (In
  Russian),  (1983).

\bibitem{lev92}
V.~I. Levenshtein, Designs as maximum codes in polynomial metric spaces,
  \emph{Acta Appl. Math.} {\bf 29}\penalty0 (1-2), \penalty0 1--82,  (1992).
\newblock ISSN 0167-8019.

\bibitem{and99}
G.~Andrews, R.~Askey, and R.~Roy, \emph{Special functions}. (Cambridge
  University Press, 1999).

\bibitem{kre77}
M.~G. Kre{\u\i}n and A.~A. Nudel{\cprime}man, \emph{The {M}arkov moment problem
  and extremal problems}. (American Mathematical Society, Providence, R.I.,
  1977).

\bibitem{mce77a}
R.~J. McEliece, E.~R. Rodemich, H.~Rumsey, and L.~R. Welch, New upper bound on
  the rate of a code via the {D}elsarte-{M}ac{W}illiams inequalities,
  \emph{IEEE Trans. Inform. Theory}. {\bf 23}\penalty0 (2), \penalty0 157--166,
   (1977).

\bibitem{kab78}
G.~A. Kabatiansky and V.~I. Levenshtein, Bounds for packings on the sphere and in
  the space, \emph{Problems of Information Transmission}. {\bf 14}\penalty0
  (1), \penalty0 3--25,  (1978).

\bibitem{sid80}
V.~M. Sidelnikov, Extremal polynomials used in bounds of code volume,
  \emph{Problemy Peredachi Informatsii}. {\bf 16}\penalty0 (3), \penalty0
  17--30,  (1980).

\bibitem{bac06}
C.~Bachoc, Linear programming bounds for codes in {G}rassmannian spaces,
  \emph{IEEE Trans. Inform. Theory}. {\bf 52}, \penalty0 2111--2126,  (2006).


\end{thebibliography}

\def\cprime{$'$} \def\cprime{$'$} \def\cprime{$'$}

\end{document}